# Photorealistic ray tracing of free-space invisibility cloaks made of uniaxial dielectrics

**Jad C. Halimeh[1,*] and Martin Wegener[2]**

[1]*Physics Department and Arnold Sommerfeld Center for Theoretical Physics, Ludwig-Maximilians-Universität München, D-80333 München, Germany*
[2]*Institut für Angewandte Physik, DFG-Center for Functional Nanostructures (CFN), and Institut für Nanotechnologie, Karlsruhe Institute of Technology (KIT), D-76128 Karlsruhe, Germany*
[*]*Jad.Halimeh@physik.lmu.de*

**Abstract:** The design rules of transformation optics generally lead to spatially inhomogeneous and anisotropic impedance-matched magneto-dielectric material distributions for, *e.g.*, free-space invisibility cloaks. Recently, simplified anisotropic non-magnetic free-space cloaks made of a locally uniaxial dielectric material (calcite) have been realized experimentally. In a two-dimensional setting and for in-plane polarized light propagating in this plane, the cloaking performance can still be perfect for light rays. However, for general views in three dimensions, various imperfections are expected. In this paper, we study two different purely dielectric uniaxial cylindrical free-space cloaks. For one, the optic axis is along the radial direction, for the other one it is along the azimuthal direction. The azimuthal uniaxial cloak has not been suggested previously to the best of our knowledge. We visualize the cloaking performance of both by calculating photorealistic images rendered by ray tracing. Following and complementing our previous ray-tracing work, we use an equation of motion directly derived from Fermat's principle. The rendered images generally exhibit significant imperfections. This includes the obvious fact that cloaking does not work at all for horizontal or for ordinary linear polarization of light. Moreover, more subtle effects occur such as viewing-angle-dependent aberrations. However, we still find amazingly good cloaking performance for the purely dielectric azimuthal uniaxial cloak.

## 1. Introduction

Transformation optics maps the geometry of a fictitious space onto actual material properties in the laboratory [1-3]. Invisibility cloaking continues to be a fascinating benchmark example to test these ideas. Generally, spatially inhomogeneous and anisotropic magneto-dielectric material distributions result. Equal magnetic and dielectric responses are required at the same time to have (i) anisotropic light propagation yet no polarization dependence of the optical response and (ii) no reflections from interfaces *via* matching of the relative optical impedance, which is given by the square root of the ratio of the magnetic permeability $\mu$ and the electric permittivity $\epsilon$. Obtaining an effective magnetic response at optical frequencies has become possible *via* three-dimensional metamaterials [4], but is necessarily connected with resonances. Hence, for passive structures, dispersion and finite losses *via* causality and the Kramers-Kronig relations are unavoidable and often unacceptable, especially in the context of macroscopic cloaking [5].

Recent experiments [6] on macroscopic broadband visible-frequency free-space invisibility cloaks made of standard uniaxial calcite have used purely dielectric anisotropic materials (calcite), *i.e.*, the magnetic permeability is set to unity everywhere. This means that the response becomes polarization dependent and not impedance-matched.

It is clear that completely neglecting the magnetic response and using uniaxial instead of biaxial materials are rather drastic *ad hoc* approximations. Apart from severely easing the experimental realization, these approximations are motivated by the fact that the behavior remains ideal for propagation of light in a two-dimensional plane and for linear polarization of light lying in that same plane. In this paper, we visualize the aberrations that occur as a result of these approximations for both polarizations of light and for more general viewing conditions in three dimensions by ray tracing. Early work on ray tracing in transformation media has been published in Refs. [7] and [8]. We investigate the paradigmatic free-space cylindrical invisibility cloak [9] with a continuously varying material distribution of a uniaxial electric permittivity tensor and unity magnetic permeability as an example. This example enables direct comparison with our previous work [10] on corresponding ideal impedance-matched magneto-dielectric cylindrical cloaks. In Sect. 2, we again use ray equations of motion for the ordinary and the extraordinary rays derived directly from Fermat's principle [11-13].

At the interfaces of the structure, double refraction occurs. While the mathematical treatment may be seen as "straightforward", we do summarize the general formulas in Sect. 3 in compact form. In the literature, usually only special cases are explicitly discussed [14-17]. Sect. 4 presents rendered images on the basis of Sect. 2 and 3. In Sect. 5 we explain the reason behind the significant difference in cloaking performance between the two variants of the uniaxial dielectric cylindrical cloak.

## 2. Ray equation of motion in non-magnetic locally uniaxial material distributions

In the geometrical optics of birefringent uniaxial dielectric materials, one has to distinguish between the paths of the ordinary and the extraordinary rays [13]. What polarization of light is ordinary and what is extraordinary will generally change at the interface between two different birefringent materials. However, if the change of the anisotropy axis $\vec{c}$ is continuous in space, the ordinary ray stays ordinary and the extraordinary ray stays extraordinary throughout even though the $\vec{c}$ axis changes. Only two different rays emerge from the overall structure (apart from further reflected contributions). We will restrict ourselves to this conceptually transparent case. Only at the interfaces of air/vacuum to the continuously changing medium, caution has to be exerted. The corresponding discussion will be given in the next section.

Inside of the structure, we can choose the local Cartesian coordinate system such that, e.g., the local $z$-axis coincides with the local $\vec{c}$ axis. In this basis, the local dielectric tensor becomes

$$\overleftrightarrow{\epsilon} = \begin{pmatrix} \epsilon_o & 0 & 0 \\ 0 & \epsilon_o & 0 \\ 0 & 0 & \epsilon_e \end{pmatrix}. \tag{1}$$

Here we have suppressed the dependence on $\vec{r}$ for better readability and we will continue doing so throughout this section. As usual, the index "o" stands for ordinary and "e" for extraordinary here and below. In addition, we assume no magnetic response at all, i.e., we have the magnetic-permeability tensor $\overleftrightarrow{\mu} = \overleftrightarrow{1}$. The ordinary and extraordinary refractive indices are then given by $n_o = \sqrt{\epsilon_o}$ and $n_e = \sqrt{\epsilon_e}$.

Let us start by discussing the propagation of the ordinary ray, for which the polarization of light, i.e., the orientation of electric-field vector $\vec{E}_o$, is perpendicular to the local $\vec{c}$ axis everywhere in the cloak. As a result, this ordinary ray merely experiences a simple refractive-index distribution equivalent to a ray in a locally isotropic yet spatially inhomogeneous medium (= "graded-index" structure). In Eq. (11) in Ref. [13] we have derived the corresponding ray equation of motion for the ordinary-ray velocity or energy velocity $\vec{v}_o$ as

$$\frac{d\vec{v}_o}{dt} = \frac{|\vec{v}_o|^2 \vec{\nabla} n_o - 2(\vec{\nabla} n_o \cdot \vec{v}_o) \vec{v}_o}{n_o}. \tag{2}$$

It should be emphasized, however, that the ordinary rays do not lead to cloaking for the cylindrical cloak to be discussed in Sect. 4.

Thus, the orthogonally polarized extraordinary rays are more important under our conditions. Their ray equation of motion is more involved though. We can closely follow the spirit of the derivation in Sect. 3 of Ref. [10], but we need to replace the expression $\vec{B}_e = \mu_0 \overleftrightarrow{\mu} \vec{H}_e = \mu_0 \overleftrightarrow{\epsilon} \vec{H}_e$ (with vacuum permeability $\mu_0$) there by an appropriate expression to again arrive at a compact form like Eq. (7) in Ref. [10]. Under the present conditions, the extraordinary magnetic field vector $\vec{B}_e$ is perpendicular to the local $\vec{c}$ axis everywhere in the cloak. Thus, its direction is invariant under multiplication with $\overleftrightarrow{\epsilon}$ (and, therefore, also $\overleftrightarrow{\epsilon}^{-1}$), just like the ordinary displacement vector $\vec{D}_o$, which is also perpendicular to $\vec{c}$. With Eq. (1), this allows us to write

$$\overleftrightarrow{\epsilon}^{-1}\vec{B}_e = \epsilon_o^{-1}\vec{B}_e = \frac{1}{n_o^2}\vec{B}_e. \tag{3}$$

Inserting Eq. (3) into the extraordinary ray velocity $\vec{v}_e$ (see Eq. (4) in Ref. [10]) gives

$$\vec{v}_e = \frac{\vec{E}_e \times \vec{H}_e}{w_e} = \frac{1}{\mu_0}\frac{\vec{E}_e \times \vec{B}_e}{w_e} = \frac{n_o^2}{\epsilon_0 \mu_0}\frac{(\overleftrightarrow{\epsilon}^{-1}\vec{D}_e) \times (\overleftrightarrow{\epsilon}^{-1}\vec{B}_e)}{w_e}, \tag{4}$$

with vacuum permittivity $\epsilon_0$ ($\neq \epsilon_o$), and the electromagnetic energy density $w_e$. Using the mathematical identity [10]

$$(\overleftrightarrow{A}\vec{a}) \times (\overleftrightarrow{A}\vec{b}) = |\overleftrightarrow{A}|\,\overleftrightarrow{A}^{-1}(\vec{a}\times\vec{b}), \tag{5}$$

where $|...|$ denotes the matrix determinant, we can connect the extraordinary ray velocity to the extraordinary wave vector of light $\vec{k}_e$ via

$$\vec{v}_e = \frac{1}{\epsilon_0 \mu_0 \omega}\overleftrightarrow{M}^{-1}\vec{k}_e, \tag{6}$$

with the angular frequency of light $\omega$ and the dimensionless auxiliary matrix

$$\overleftrightarrow{M} = \frac{1}{n_o^2}|\overleftrightarrow{\epsilon}|\,\overleftrightarrow{\epsilon}^{-1} = n_o^2 n_e^2\,\overleftrightarrow{\epsilon}^{-1}. \tag{7}$$

In the last step in Eq. (7) we have inserted the determinant $|\overleftrightarrow{\epsilon}| = n_o^4 n_e^2$ (see Eq. (1)). Note that all quantities in Eq. (7) are generally still dependent on the coordinate vector $\vec{r}$. Inserting Eq. (6) and Eq. (7) into Fermat's principle [10] and working out the Euler-Lagrange equations [10] leads us to the final extraordinary ray equation of motion (see Eq. (11) in Ref. [10]) in compact form analogous to Newton's second law [10]

$$\frac{d\vec{v}_e}{dt} = \frac{\overleftrightarrow{M}^{-1}}{2}\left[\vec{v}_e(\vec{\nabla}\overleftrightarrow{M})\vec{v}_e - 2\left((\vec{\nabla}\overleftrightarrow{M})\cdot\vec{v}_e\right)\vec{v}_e\right]. \tag{8}$$

Obviously, this ray equation of motion is of the same form as that derived for impedance-matched anisotropic magneto-dielectrics Eq. (11) in Ref. [10], but the auxiliary matrix $\overleftrightarrow{M}$ for non-magnetic uniaxial dielectrics in Eq. (7) is generally different from Eq. (8) in Ref. [10].

## 3. Interface of the structure to air/vacuum

Arbitrarily polarized incident light impinging from the outside of the cloak (*i.e.*, from vacuum or air) will be double refracted at the cloak's interface into an ordinary and an extraordinary ray, which will then propagate inside the cloak. Thus, we first need to determine the amplitudes and directions of these two rays before being able to apply the results of the previous section.

In the following, the general case is considered where the optic axis $\vec{c}$ is not necessarily in the plane of incidence, which is the case in the azimuthal uniaxial cloak discussed in the next section. Two planes are defined: the ordinary main section, formed by the optic axis and the ordinary refracted wave vector, and the extraordinary main section, formed by the optic axis and the extraordinary refracted wave vector. The ordinary wave has its polarization vector orthogonal to the ordinary main section, while the extraordinary wave has its polarization vector in the extraordinary main section. The ordinary wave thus sees a refractive index equal to $n_o$ whereas the refractive index $n$ that the extraordinary wave sees is given by [15]

$$n(\phi_e) = \frac{n_o n_e}{\sqrt{n_o^2 \cos^2(\phi_e) + n_e^2 \sin^2(\phi_e)}}, \tag{9}$$

where $\phi_e$ is the angle the extraordinary polarization vector includes with the unit optic axis $\vec{c}$. If $\varphi_e$ is the angle of refraction of the extraordinary wave, one determines that

$$\phi_e = \arcsin\left(\left(\hat{\vec{c}}\cdot\hat{\vec{t}}\right)\sin(\varphi_e) - \left(\hat{\vec{c}}\cdot\hat{\vec{n}}\right)\cos(\varphi_e)\right), \tag{10}$$

where $\hat{\vec{n}}$ is the unit normal vector and $\hat{\vec{t}}$ is the unit tangential vector in the plane of incidence (one can see that in the case of the radial uniaxial dielectric cloak, where the optic axis is parallel to the unit normal, we have, e.g., $\phi_e = \pi/2 - \varphi_e$). As such, $n = n(\varphi_e)$ can now be expressed as a function of $\varphi_e$, which enters into the equation of phase matching

$$n_i \sin(\varphi_i) = n_o \sin(\varphi_o) = n(\varphi_e)\sin(\varphi_e), \tag{11}$$

where $n_i=1$ is the refractive index of air/vacuum, $\varphi_i$ is the angle of incidence, and $\varphi_o$ is the angle of ordinary refraction. From Eq. (11), it is straightforward to calculate $\varphi_o$. To determine $\varphi_e$, one must solve the implicit Eq. (11) from which two solutions emerge, with only one of them acceptable being in $[0, \pi/2]$. As such, the directions of the refracted waves are determined. The amplitudes of the wave vectors are proportional to their respective refractive indices due to the condition of phase matching, and therefore the ordinary wave vector will carry a magnitude $k_o = n_o \omega/c_0$ (with the vacuum speed of light $c_0$) and that of the extraordinary wave a magnitude of $k_e = n(\varphi_e)\omega/c_0$.

Note that the above procedure allows determining the directions of all fields since the directions of the wave vectors are now known and so are the geometric conditions for their fields. For example, taking into account the extraordinary wave, it is known that its polarization $\vec{D}_e$ must lie in the extraordinary main section. Moreover, a consequence of Maxwell's equations is that $\vec{D}_e$ is perpendicular to $\vec{k}_e$, which exactly determines its direction within that plane. Moreover, another consequence of Maxwell's equations is that $\vec{H}_e \perp \vec{D}_e$, and, as the medium is non-magnetic, this means that $\vec{B}_e \perp \vec{D}_e$. Since $\vec{B}_e$ is also perpendicular to $\vec{k}_e$ due to Maxwell's equations, $\vec{B}_e$ is then perpendicular to the extraordinary main section, thus exactly determining its direction. With the directions of all fields determined, the boundary conditions of continuity [14] then allow us to determine their exact magnitudes. The ray-velocity vector can then be determined as per Eq. (4) or Eq. (6). The intensity transmission coefficients $T_o$ and $T_e$ for the ordinary and extraordinary refracted rays, respectively, are given by [16]

$$T_o = \left|\frac{\hat{\vec{n}}\cdot(\vec{E}_o \times \vec{H}_o)}{\hat{\vec{n}}\cdot(\vec{E}_i \times \vec{H}_i)}\right| \quad \text{and} \quad T_e = \left|\frac{\hat{\vec{n}}\cdot(\vec{E}_e \times \vec{H}_e)}{\hat{\vec{n}}\cdot(\vec{E}_i \times \vec{H}_i)}\right|, \tag{12}$$

where the index "i" stands for the incident quantity.

Although mathematically similar, it is worth briefly mentioning that in the case of an interface from a uniaxial dielectric crystal to air/vacuum, there will in general be double reflections within the crystal, into an ordinary reflected wave and an extraordinary reflected wave, irrespective of whether the incident wave is itself ordinary or extraordinary. The directions and amplitudes of all relevant fields are determined just as in the case of the air/vacuum-cloak interface, except here there are two reflections (in cloak) and one refraction (into air), instead of one reflection (in air) and two refractions (in cloak) at the interface. Moreover, here it is the reflected wave vectors (ordinary and extraordinary) that, along with the optic axis, define the main sections.

The intensity reflection coefficients are unity minus the transmission coefficient.

## 4. Numerical results

In cylindrical coordinates, the ideal magneto-dielectric parameters for the cylindrical cloak [9] with inner radius $a$ and outer radius $b$ are given by [9]

$$\epsilon_r = \mu_r = \frac{r-a}{r} \; ; \; \epsilon_\Theta = \mu_\Theta = \frac{r}{r-a} \; ; \; \epsilon_z = \mu_z = \left(\frac{b}{b-a}\right)^2 \frac{r-a}{r}. \tag{13}$$

(Unfortunately, Eq. (12) in Ref. [10] contained a typo.) As discussed in the introduction, we wish to eliminate the magnetic response, *i.e.*, we want to use the reduced magnetic parameters $\mu'_r = \mu'_\theta = \mu'_z = 1$. At the same time, we want to keep the cloaking performance exact – at least for light propagating in the $r\theta$-plane with an electric-field vector lying in that plane (hence a magnetic-field vector along $z$). To achieve this goal, the refractive-index distribution for the reduced parameters has to be the same as for the ideal parameters Eq. (2), *i.e.*, the conditions $\epsilon_r \mu_z = \epsilon'_r \mu'_z$ and $\epsilon_\theta \mu_z = \epsilon'_\theta \mu'_z$ need to be fulfilled. The same spirit has previously been followed in Ref. [18]. Otherwise, the propagation time of light would be different with and without cloak (compare Ref. [10]) and the structure may be a ray cloak but just cannot be a wave cloak [10]. For example, the 3D free-space structure realized in Ref. [6] is a ray cloak but not a wave cloak, while the 3D carpet cloak in Ref. [19] has been shown to work for light waves and rays. For light propagating in the $r\theta$-plane with polarization in that plane, the reduced $\epsilon'_z$ component does not enter at all. Thus, it is not yet determined. This leaves us with two non-equivalent choices.

We can choose the axial component to be identical to the azimuthal component. With Eq. (13), this leads us to the reduced dielectric parameters

$$\epsilon'_r = \left(\frac{r-a}{r}\right)^2 \left(\frac{b}{b-a}\right)^2 \; ; \; \epsilon'_\theta = \left(\frac{b}{b-a}\right)^2 \; ; \; \epsilon'_z = \left(\frac{b}{b-a}\right)^2. \tag{14}$$

We shall call the cloak according to Eq. (14) the radial uniaxial cloak as its optic axis is obviously along the radial direction. The other choice is the azimuthal uniaxial cloak, for which we get

$$\epsilon'_r = \left(\frac{r-a}{r}\right)^2 \left(\frac{b}{b-a}\right)^2 \; ; \; \epsilon'_\theta = \left(\frac{b}{b-a}\right)^2 \; ; \; \epsilon'_z = \left(\frac{r-a}{r}\right)^2 \left(\frac{b}{b-a}\right)^2. \tag{15}$$

To allow for direct comparison with our previous results on the ideal cylindrical cloak [10], we use the exact same scenery (see Fig. 3 in Ref. [10]) and the identical parameters as in Ref. [10], *i.e.*, $a = 5$ cm and $b = 10$ cm. Rendered ray-tracing images as well as time-of-flight (TOF) difference images for the ideal cylindrical cloak have been shown in Fig. 4 in Ref. [10].

In the numerical implementation, we account for first-order reflections but omit multiple-order reflections. This is justified because even the first-order reflected components (except for those from the inner metal cylinder) turn out to be barely visible in the rendered images.

We start with unpolarized light rays emitted from the sources. Sect. 3 has given closed exact expressions for treating double refraction at the entrance interface from air/vacuum with $\epsilon = \mu = 1$ to the cloak. We then solve Eq. (2) for the ordinary ray and Eq. (8) for the extraordinary ray inside of the cloak numerically using a standard fourth-order Runge-Kutta approach. Regarding reflection of light at the inner metal cylinder inside of the cloak, one must generally be careful because of possible bi-reflections within birefringent media [16]. However, for the radial uniaxial cloak with Eq. (14), the local $\vec{c}$ axis (*i.e.*, the radial direction) coincides with the metal surface normal and, thus, we get only a single reflected ray, the reflection angle of which simply equals the incidence angle for the ordinary as well as for the extraordinary ray. For the case of the radial uniaxial cloak below, all ordinary rays reach the inner metal cylinder while no extraordinary rays do. For the azimuthal uniaxial cloak, the rays turn out to never reach the inner metal cylinder for our conditions. Finally, refraction at the exit interface towards air/vacuum is again treated analytically according to Sect. 3.

For the radial uniaxial cloak according to Eq. (14), the ordinary refractive index is constant within the cloak ($n_o = 2$ for the chosen parameters). Thus, the right-hand side of Eq. (2) is zero everywhere inside of the cloak. This means that ordinary rays propagate along straight lines within the cloak. The ordinary rays are, however, refracted at the cloak interfaces like for an ordinary cylindrical lens. The right-hand side of Eq. (2) is nonzero for the azimuthal uniaxial cloak according to Eq. (15), *i.e.*, ordinary rays are curved within the cloak.

Let us mention in passing that we have also modified and programmed the Hamiltonian ray tracing as in Ref. [7] for the present conditions. Situations for which perfect cloaking is expected have been used as test examples. To obtain the same relative accuracy of, *e.g.*, $10^{-8}$ with the Runge-Kutta method, the Hamiltonian approach needed more than one order of magnitude longer computation times than ours.

Rendered images for the non-magnetic, radial uniaxial dielectric, cylindrical cloak according to Eq. (14) are depicted in Fig. 1. Panel (a) shows the reference image without cylinder, (b) with metal cylinder of radius a, and (c) with metal cylinder and with cloak around it – all for unpolarized light. Clearly, (c) is the admixture of two rather different images for the two orthogonal linear polarizations. In panel (d), the virtual observer wears polarization goggles transmitting only vertical polarization and only horizontal polarization in (e). Note, however, that the polarization basis of the goggles is generally different from the polarization basis of an incoming ray. Hence, both panels (d) and (e) are admixtures of ordinary and extraordinary rays emerging from the cloak.

For a normal observer, it is fair to say that the device simply does not work as an invisibility cloak (see Fig. 1(c)). In contrast, for selecting only vertical linear polarization in the image, the amplitude image in Fig. 1(d) exhibits quite reasonable cloaking in the center of the image and along the vertical direction. The cloaking performance deteriorates when departing from these cases, *e.g.*, by rotating the observation axis around the center of the image, where we observe an adverse *rotational effect* on the cloaking quality. Recall that the horizontal field-of-view (FOV) in Fig. 1 is 50° (see Fig. 3 in Ref. [10]), emulating a human focal FOV. For horizontal polarization, the "cloak" exactly acts like an isotropic dielectric around a reflective metal cylinder and, hence, represents a bad cylindrical lens with a curved reflector inside (see Fig. 1(e)). As to be expected from the design, "cloaking" is so bad for horizontal polarization that it deserves no further discussion.

Due to the design of the cloak described above, the time-of-flight (TOF) images [10] for vertical analyzer orientation, in the center of the image and along the vertical direction, and for the case of cloak and cylinder are expected to be perfectly identical to the case of no cylinder (not depicted).

The cloaking performance of the radial uniaxial dielectric cloak according to Eq. (14) is very good in the center of the image along the vertical direction in Fig. 1(d). However, as soon as one rotates the observation axis, huge distortions occur. Thus, the overall performance of this cloak, the design of which closely follows Ref. [18], is quite disappointing in three dimensions and even in two dimensions away from the center vertical plane.

Fig. 2 shows the same scenario and the same cases as in Fig. 1, but for the azimuthal uniaxial dielectric cloak according to Eq. (15). To the best of our knowledge, this design has not been discussed before. Again, as expected by the design, horizontal orientation of the polarization goggles in Fig. 2(e) is so bad that it deserves no further discussion. In sharp contrast, the performance is rather good for vertical polarization in Fig. 2(d). This is expected from the cloak design as described above for the middle of the image along the vertical direction. The good performance is much less obvious for other viewing directions and could only be revealed by our rendering. An intuitive explanation as to why the azimuthal uniaxial cloak performs so much better than the radial one will be given in the next section.

In a hypothetical TOF experiment [10] using a short pulse incident onto the scenery, one would, however, observe satellites in addition to the main transmitted pulse, by which the cloak could likely be revealed.

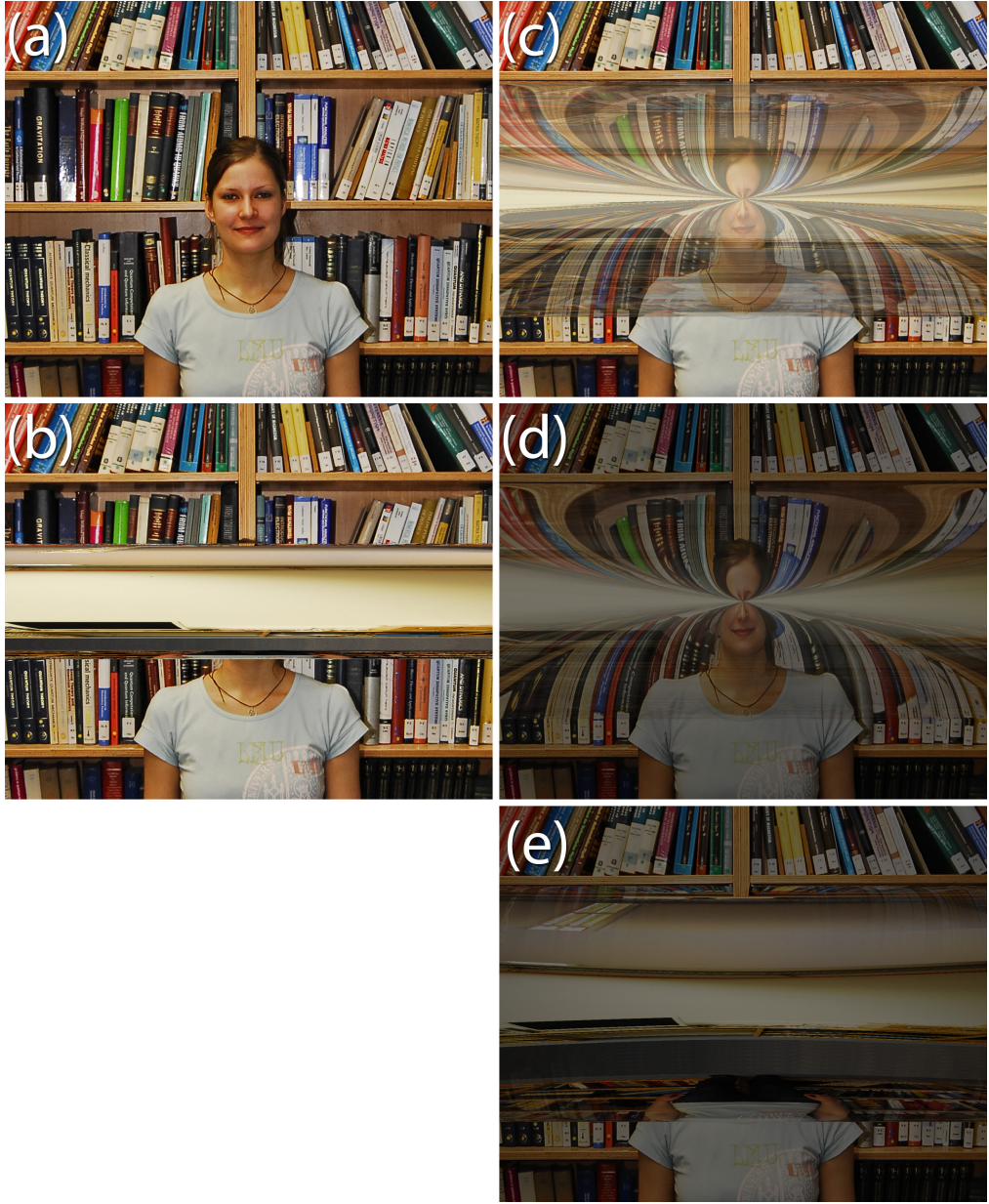

Fig. 1. Photorealistic images for the non-magnetic radial uniaxial dielectric cloak according to Eq. (14) rendered by ray tracing using Eq. (2) and Eq. (8). The scenery and the input raw images are defined in Fig. 3 in Ref. [10]. (a) Reference image without metal cylinder, (b) with metal cylinder in front of the model's head but without cloak, (c) with metal cylinder and with radial uniaxial dielectric cloak and for unpolarized light, (d) as (c) but for vertical orientation of the linear polarizer in front of the virtual camera, and (e) as (c) but for horizontal polarization.

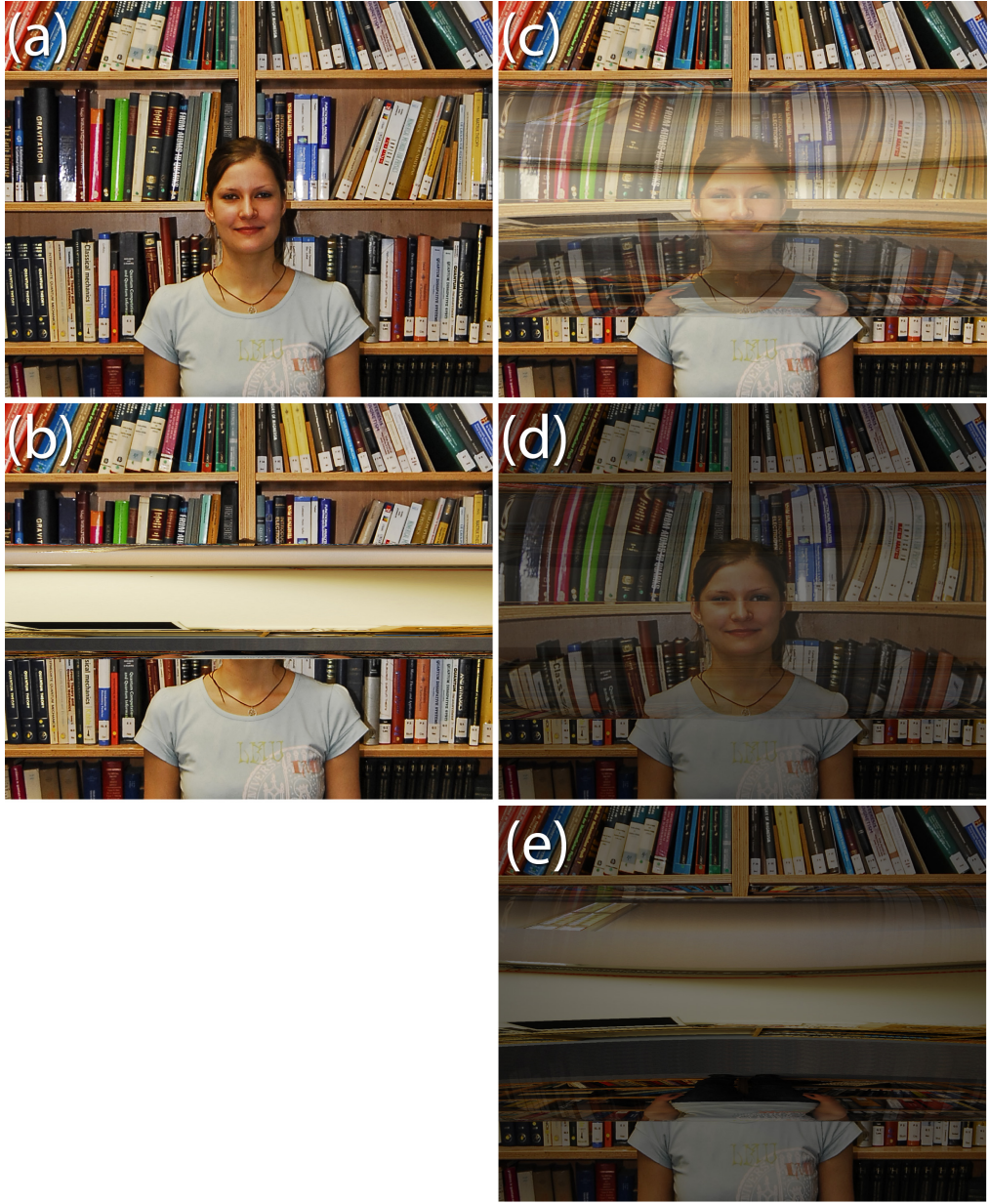

Fig. 2. As Fig. 1, but for the non-magnetic azimuthal uniaxial dielectric cloak according to Eq. (15) rather than the radial uniaxial cloak according to Eq. (14). Comparison of panels (d) of Fig. 1 and Fig. 2 shows a far superior performance of the azimuthal uniaxial dielectric cloak compared to the radial uniaxial dielectric cloak. In panels (d), the linear polarizer in front of the virtual camera is vertically orientated. This different cloaking performance is explained in Sect. 5 and illustrated in Fig. 3.

## 5. Intuitive explanation

The rendered images of the preceding section have shown that the azimuthal uniaxial purely dielectric cloak performs much better than its radial counterpart. Here, we aim at providing an intuitive explanation for this difference.

In Fig. 3, we consider for each of the two cloaks two incident wave vectors, namely $\vec{k}^v$ in the vertical plane and $\vec{k}^h$ in the horizontal plane, both of which can be assumed to emanate from, *e.g.*, the viewing camera, and examine what happens only to their extraordinary refracted wave vectors $\vec{k}_e^v$ and $\vec{k}_e^h$, respectively. Noting that the polarization vector of the refracted extraordinary wave vector must lie in the extraordinary main section, we see that for both cloaks the extraordinary polarization vector $\vec{D}_e^v$ is indeed in the vertical plane when the incident wave vector is along $\vec{k}^v$, and thus the cloaking should be perfect in this case since the cloaks are designed to cloak perfectly for light rays that, along with their polarization vectors, lie parallel to the vertical plane.

However, if the incident wave vector is parallel to $\vec{k}^h$, the situation is drastically different. In the case of the radial uniaxial cloak, where the optic axis is parallel to the surface normal, we see that the extraordinary main section is identical to the horizontal plane itself in this case, which is also the plane of incidence. This means that the extraordinary polarization vector $\vec{D}_e^h$ lies completely in the horizontal plane, and thus the cloaking is expected to be significantly deteriorated in this case. One can envisage that, as the observation axis is rotated around the center of the image, incident rays onto the cloak will partially refract into extraordinary rays that will behave in a manner that is in between these two extremes. This leads to what we term the rotational effect present in the cloaking behavior of the radial uniaxial cloak. In contrast, in the case of the azimuthal uniaxial cloak, the rotational effect is completely absent. Here, $\vec{D}_e^h$ will be parallel to the vertical plane since the optic axis here is perpendicular to the horizontal plane leading the extraordinary main section to be perpendicular to the plane of incidence (horizontal plane). As such, $\vec{k}^h$ will experience a refractive-index distribution which is much more favorable for cloaking. The only reason the cloaking is imperfect here (albeit still considerably better than in the case of the radial uniaxial cloak) is due to the fact that $\vec{k}^h$ itself is not parallel to the vertical plane.

## 6. Conclusion

We have derived Newton's-law-like ray equations of motion for non-magnetic continuously varying anisotropic locally uniaxial dielectric material distributions directly from Fermat's principle. This complements our previous corresponding work on graded-index locally isotropic dielectrics [13] and on locally anisotropic impedance-matched magneto-dielectric material distributions [10] – both of which show no birefringence.

We have used this approach to render photorealistic images of cylindrical free-space invisibility cloaks made of locally anisotropic non-magnetic dielectrics, using home-built, dedicated ray-tracing software. Two cases have been discussed: the radial and the azimuthal uniaxial cloaks. The radial uniaxial cloak performs rather badly overall – even for its design polarization.

In sharp contrast, the purely dielectric azimuthal uniaxial cloak, which has not been suggested previously, exhibits surprisingly good cloaking in three dimensions for one linear polarization of light. In particular, the performance goes well beyond that of recent impressive experiments [6] in that it leads to the correct time-of-flight ("wave cloak"), at least in the center of the image, whereas the piece-wise homogeneous calcite polygonal structure in Ref. [6] is expected to exhibit huge time-of-flight differences compared to empty space and is hence only a "ray cloak". Moreover, in contrast to Ref. [6], our cloak and the resulting performance are rotationally invariant, *i.e.*, the performance does not depend on the viewing direction. The price to pay in our case is that two components of the local permittivity tensor assume positive values below one, *i.e.*, we need superluminal light propagation.

It is often believed that such values of the permittivity tensor components $0 \leq \epsilon_{ij} \leq 1$ are only possible in a narrow frequency interval (*e.g.*, at frequencies above the resonance frequency of a pronounced resonance) and that causality demands that they are accompanied by serious losses. Both aspects have to be treated with extreme caution. Recent experiments on metamaterials composed of active non-Foster elements operating at MHz and GHz frequencies [20,21] have shown rather constant superluminal electromagnetic phase and group velocities from 2 MHz to 40 MHz frequency [21] (more than four octaves bandwidth). The frequency dependence occurring at yet higher frequencies guarantees the fundamental consistency with causality and with the laws of special relativity [20,21].

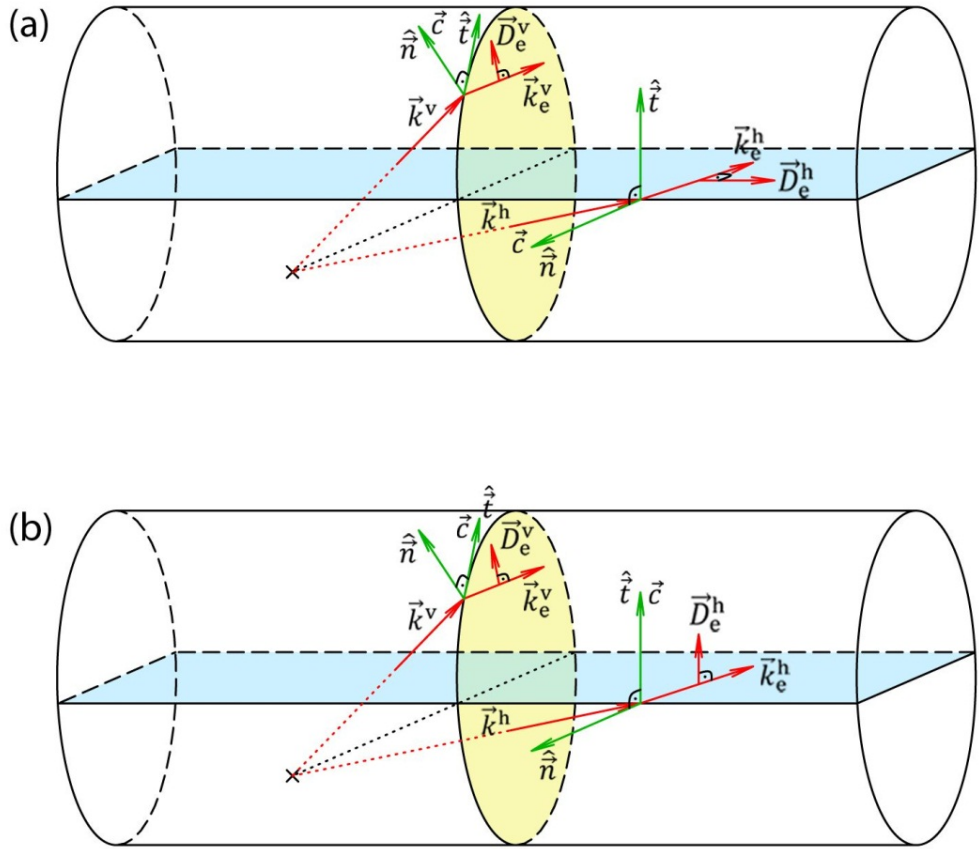

Fig. 3. Illustration explaining the different cloaking behavior found in Fig. 1 and Fig. 2. (a) Radial uniaxial cylindrical cloak and (b) azimuthal uniaxial cylindrical cloak. Extraordinary light in the vertical plane carries an extraordinary polarization vector that is parallel to the vertical plane in both cases, thus providing perfect cloaking as expected. However, for extraordinary light in the horizontal plane, the extraordinary polarization vector lies in the horizontal plane in the case of the radial uniaxial cloak, leading to bad cloaking quality. In contrast, for the case of the azimuthal uniaxial cloak, the extraordinary polarization is still parallel to the vertical plane, which generally leads to good cloaking behavior.

**Acknowledgements**

We thank our model, Tanja Rosentreter, and the photographer, Vincent Sprenger (TU München), for help with the photographs used as input for the photorealistic ray-tracing calculations. We are grateful to Maximilian Papp for his artistic help with Fig. 3 and the layout of all Figures. J.C.H. thanks Michael Kay (LMU München) for fruitful discussions. We acknowledge the support of the Arnold Sommerfeld Center (LMU München), which allowed us to use their computer facilities for our rather CPU-time-consuming numerical ray-tracing calculations. J.C.H. acknowledges financial support from the Deutsche Forschungsgemeinschaft (DFG) through FOR801 and by the Excellence Cluster "Nanosystems Inititiative Munich (NIM)". M.W. acknowledges support by the DFG, the State of Baden-Württemberg, and the Karlsruhe Institute of Technology (KIT) through the DFG Center for Functional Nanostructures (CFN) within subproject A1.5.